\documentclass[fleqn,usenatbib,useAMS]{mnras}

\usepackage{newtxtext,newtxmath}

\usepackage[T1]{fontenc}

\DeclareRobustCommand{\VAN}[3]{#2}
\let\VANthebibliography\thebibliography
\def\thebibliography{\DeclareRobustCommand{\VAN}[3]{##3}\VANthebibliography}

\usepackage{graphicx}	
\usepackage{amsmath}	
\usepackage{xspace}

\newcommand{\fgl}{1FGL J1018.6-5856\xspace}
\newcommand{\psrb}{PSR\,B1259-63\xspace}
\newcommand{\psrj}{PSR\,J2032+4127\xspace}
\newcommand{\msun}{\ensuremath{{\rm M}_\odot}}
\newcommand{\cha}{Chandra\xspace}
\newcommand{\xmm}{XMM-Newton\xspace}
\newcommand{\xrt}{Swift/XRT\xspace}

\title[Improved binary solution for 1FGL J1018.6–5856]{Improved binary solution for the gamma-ray binary 1FGL J1018.6–5856\thanks{Based on observations made with the Southern African Large Telescope (SALT) under
programmes 2018-1-MLT-001 and 2020-1-DDT-003 (PI: B van Soelen)}}

\author[B.~van Soelen, S.~Mc. Keague, D.~Malyshev et al.]{
B.~van Soelen,$^{1}$\thanks{E-mail: vansoelenb@ufs.ac.za}
S.~Mc~Keague,$^{2}$
D.~Malyshev,$^{3}$
M.~Chernyakova,$^{2,4}$
N.~Komin,$^{5}$
N.~Matchett,$^{1}$
\newauthor
and I.M.~Monageng$^{6,7}$
\\
$^{1}$Department of Physics, University of the Free State, PO Box 339, Bloemfontein 9300, South Africa\\
$^{2}$School of Physical Sciences and Centre for Astrophysics \& Relativity, Dublin City University, Glasnevin, D09 W6Y4, Ireland\\
$^{3}$Institut f{\"u}r Astronomie und Astrophysik T{\"u}bingen, Universit{\"a}t T{\"u}bingen, Sand 1, D-72076 T{\"u}bingen, Germany\\
$^{4}$Dublin Institute for Advanced Studies, 31 Fitzwilliam Place, Dublin 2\\
$^{5}$School of Physics, University of the Witwatersrand, 1 Jan Smuts Avenue, Braamfontein, Johannesburg, 2050 South Africa\\
$^{6}$ South African Astronomical Observatory, PO Box 9, Observatory, 7935, Cape Town, South Africa\\ 
$^{7}$ Department of Astronomy, University of Cape Town, Private Bag X3, Rondebosch 7701, South Africa\\
}

\date{Accepted XXX. Received YYY; in original form ZZZ}

\pubyear{2022}

\begin{document}
\label{firstpage}
\pagerange{\pageref{firstpage}--\pageref{lastpage}}
\maketitle


\begin{abstract}
The gamma-ray binary \fgl\ consists of an O6V((f)) type star and an unknown compact object, and shows orbitally modulated emission from radio to very high energy gamma rays. The X-ray light curve shows a maximum around the same phase as the GeV emission, but also a secondary maximum between phases $\phi=0.2 - 0.6$.  A clear solution to the binary system is important for understanding the emission mechanisms occurring within the system. In order to improve on the existing binary solution, we undertook radial velocity measurements of the optical companion using the Southern African Large Telescope, as well as analysed publicly available X-ray and GeV gamma-ray data.  A search for periodicity in {\it Fermi}-LAT data found an orbital period of  $P=16.5507\pm0.0004$\,d. The best fit solution to the radial velocities, held at this new period, finds the system to be more eccentric than previous observations, $e=0.531 \pm 0.033$ with a longitude of periastron of $151.2 \pm 5.1$\degr, and a larger mass function $f =  0.00432\pm 0.00077$\,M$_\odot$. We propose that the peaks in the X-ray and gamma-ray light curves around phase 0 are due to the observation of the confined shock formed between the pulsar and stellar wind pointing towards the observer. The secondary increase or strong rapid variations of the X-ray flux at phases 0.25-0.75 is due to the interaction of multiple randomly oriented stellar wind clumps/pulsar wind
interactions around apastron.
\end{abstract}




\section{Introduction}

Gamma-ray binaries are a small, but growing subset of high mass binary systems that produce the majority of their gamma-ray emission above 1\,MeV, with less than 10 sources identified to date \citep[see e.g.][]{cta_binary19,2013A&ARv..21...64D}. All gamma-ray binaries consist of a compact object, either a neutron star or black hole, in orbit around an O/Oe or B/Be type star. For only two systems, namely \psrb\ and \psrj, is the nature of the compact known, identified as young, rapidly rotating pulsars \citep{johnston92,camilo09}. In these systems the emission originates from the shock that forms between the pulsar and stellar winds.  In the other binary systems the nature of the compact object has not been identified and the possibility of a microquasar like scenario must also be considered \citep[see e.g.][]{massi13}. 

The discovery of the gamma-ray binary \fgl\ was reported in \cite{Fermi_1FGL1018}, finding modulated GeV gamma-ray, radio and optical emission, with an orbital period of $16.58\pm0.02$\,d. Interestingly, while the radio and GeV light curves both showed a single maximum (although the radio is in anti-correlation with the GeV),  the X-ray light curve showed two maxima: a broad maximum peaking near phase $\phi=0.4$ and a narrower, brighter maximum at $\phi\approx0$, the same phase as the maximum in the GeV light curve.\footnote{The reference time for $\phi=0$ has been defined differently by different authors. The discovery paper established the reference time $T_0 =  55403.3$ (MJD) based on the peak in the {\it Fermi}-LAT light curve. However, \cite{strader15} and subsequently \citet{monageng17} adopted $T_0 = 57244.36$, based on the time of the ascending node. In this paper we will use the reference time as established in the discovery paper.} This was also seen in subsequent X-ray observations \citep[e.g.][]{an13}. The system has subsequently been detected and shown to be variable at very high energy gamma-rays by the H.E.S.S.\ telescope \citep{2012A&A...541A...5H,HESS_1FGL1018}. 
The most updated orbital period for the system, determined from X-ray observations, is $16.544\pm 0.008$\,d \citep{an15}.

The optical companion is classified as a O6V((f)) type star \citep{Fermi_1FGL1018} (USNO-B1.0 0310-00164673; ${\rm RA}=10^\rmn{h} 18^\rmn{m} 55\fs5874834232$ ${\rm Dec}= -58\degr 56\arcmin 45 \farcs 974782738$). The distance to the source was found to be $6.4^{+1.7}_{-0.7}$\,kpc by \cite{marcote18}, but we note that the {\it Gaia Early Release Data Release 3} gives an updated parallax of $0.2273 \pm 0.0102$ mas, which would correspond to a distance of $4.4 \pm 0.2$\,kpc \citep
{gaia_mission_16, gaia_erdr3}, consistent with earlier measurements based on the photometry by \cite{napoli11}.

The solution for the orbital parameters of the system has been considered by a number of authors, based on radial velocity measurements of the optical companion  \citep{waisberg15,strader15,monageng17}.  A model dependent orbital solution was also proposed by \cite{an17}, who suggested the two maxima in the X-ray light curve could be explained by an eccentric orbit ($e=0.35$) with the first maximum at periastron ($\phi=0.4$),  and Doppler boosted emission producing the second maximum near inferior conjunction, which they placed at $\phi=0$. However, this solution is at odds with the binary solution found by \cite{monageng17}, who found inferior conjunction and periastron at a similar phase ($\phi \sim 0$). The details of the binary geometry are, therefore, important for understanding the emission mechanisms occurring within the system. 

In this paper we present new high resolution optical spectroscopic observations undertaken with the Southern African Large Telescope (SALT), to improve the binary solution for \fgl. Since the phase of periastron is also dependent on the adopted orbital period, we have re-investigated the orbital period, by searching for periodicity in the approximately 13 years of {\it Fermi}-LAT observations. We have also analysed all publicly available X-ray data to investigate the double peak structure in the light curve.

\section{Observations}

\subsection{SALT observations}
\fgl\ was successfully observed 23 times, between 2018 May 2 and 2020 June 14, using the High Resolution Spectrograph \citep[HRS; ][]{bramall10,bramall12,crause14}, in Low Resolution (LR; R=14\,000) mode. Each observation consisted of two camera exposures of 1\,675 seconds. The spectra were extracted, wavelength calibrated, and flat corrected using the SALT/HRS pipeline \citep[described in][]{kniazev16}. Cosmic ray cleaning was run on each order, for both the target and sky fibre, before performing sky subtraction, and merging and continuum correcting the spectra using {\sc iraf/pyraf}. The barycentric  velocity correction of each observation was determined using {\sc astropy} \citep{astropy18} and the observation was corrected to the barycentre using {\sc dopcor}. Nightly exposures were then averaged together. 

\begin{figure}
    \centering
    \includegraphics[width=\columnwidth]{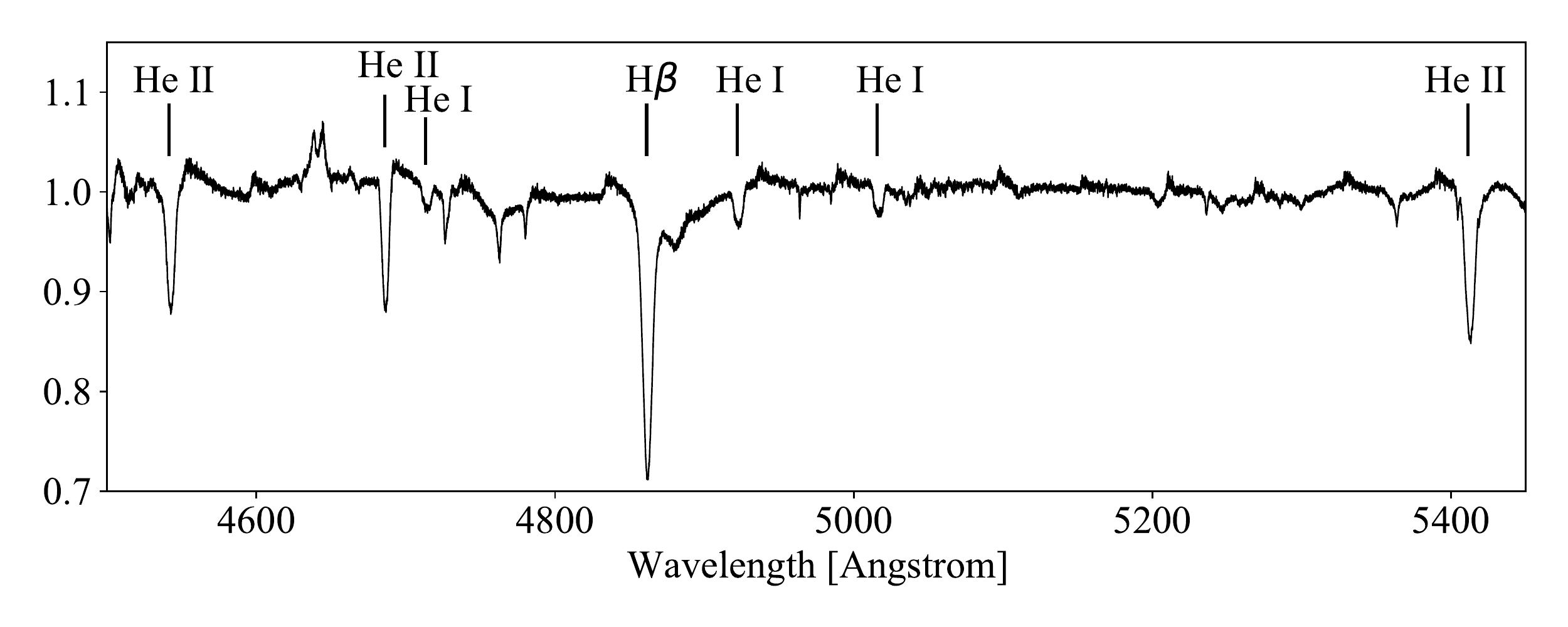}
    \caption{The average normalized spectrum for \fgl. Prominent H and He lines are indicated.}
    \label{fig:template_spectrum}
\end{figure}

The radial velocity of each observation was determined by cross-correlating it, between 4500--5450\,\AA{}, against a template constructed from the observations,  using the {\sc rvsao/xcsao} package \citep{kurtz98}. The template was constructed by determining the velocity difference between each observation, and the average of all observations through cross-correlation, then correcting each observation for this velocity. These velocity corrected observations were then averaged together to create the template against which each observation was compared. When the final radial velocity was measured, the specific observation being measured was excluded from the template. This procedure is similar to that followed in, for example, \citet{foellmi03}, \citet{monageng17}, and \citet{vansoelen19}. The average template, for the 4500--5450\,\AA{} wavelength range, is shown in Fig.~\ref{fig:template_spectrum}.  The wavelength range used for the cross-correlation includes the H\,$\beta$ line, three He\,{\sc i} and four He\,{\sc II} lines. Some regions of poorer order merging were masked in the final cross-correction, but tests showed that this did not have a significant influence on the final velocities.  The measured radial velocities, relative to the template are listed in Table~\ref{tab:radial_velocities}.

\begin{table}
    \centering
        \caption{Velocities of each observation relative to the template. The average template was measured to have a velocity of $56.8 \pm 13.1$  km\,s$^{-1}$. }
    \label{tab:radial_velocities}
    \begin{tabular}{lr} \hline 
    MJD & Velocity (km\,s$^{-1}$) \\ \hline
58240.8808  & $  1.7  \pm  1.66  $ \\
58241.8771  & $  2.05  \pm  0.88  $ \\
58242.8670  & $  7.26  \pm  0.82  $ \\
58244.8634  & $  8.86  \pm  1.16  $ \\
58264.7989  & $  6.23  \pm  1.8  $ \\
58265.7946  & $  4.6  \pm  1.22  $ \\
58281.7516  & $  2.26  \pm  1.26  $ \\
58282.7513  & $  -2.7  \pm  1.12  $ \\
58472.0387  & $  0.71  \pm  1.03  $ \\
58477.0082  & $  10.06  \pm  1.52  $ \\
58478.0128  & $  4.82  \pm  1.09  $ \\
58506.9306  & $  3.94  \pm  1.02  $ \\
58507.9254  & $  1.8  \pm  1.46  $ \\
58519.8826  & $  -3.44  \pm  0.91  $ \\
58527.8820  & $  9.43  \pm  2.08  $ \\
58531.8651  & $  -7.88  \pm  1.26  $ \\
58532.8652  & $  -26.9  \pm  1.38  $ \\
58557.7921  & $  5.7  \pm  1.15  $ \\
58558.7900  & $  5.08  \pm  1.03  $ \\
58560.7902  & $  8.23  \pm  0.97  $ \\
58864.9718  & $  -14.87  \pm  2.38  $ \\
58996.8050  & $  -22.62  \pm  0.85  $ \\
59014.7336  & $  -12.1  \pm  0.73  $ \\\hline
    \end{tabular}
\end{table}

The orbital parameters were found by fitting the radial velocities using the {\sc helio\_rv} package, part of the IDL Astronomy Library \citep{landsman}.\footnote{https://idlastro.gsfc.nasa.gov/} In order to better account for any additional systematic errors, all reported errors found using this package are determined by scaling the error of the radial velocity measurements to find a reduced $\chi^2$ value of 1 \citep{lampton76}. 

\subsection{X-ray observations}
\subsubsection{Chandra}
\fgl was observed by \cha from 2010-2013. We  analysed the three available observations (ObsId:11831,14657,16560) with the \texttt{CIAO} v.4.13 (CALDB v.4.9.5) software. The data were reprocessed with the \texttt{chandra\_repro} utility, source and background spectra with corresponding RMFs and ARFs were extracted with the \texttt{specextract} tool. The source spectrum was extracted from a $30''$-radius circle region centred at the position of \fgl, while the background was obtained from a circle region located in a source-free vicinity.
We modelled the spectra with the help of the \texttt{Xspec} software and found that the spectra could be well described by an absorbed power-law model (\texttt{cflux*TbAbs*powerlaw}). The best-fit parameters of the model as well as the basic details of the observations are summarised in Table~\ref{tab:goodxrays}.

\begin{table*}
    \centering
        \caption{High-quality observations of \fgl taken by \cha and \xmm. The table summarises Observational Id, date, exposure, flux (0.3-10\,keV) as well as best-fit parameters of the spectra with absorbed power-law model -- the power-law index and hydrogen column density $n_H$.}
    \label{tab:goodxrays}
    \begin{tabular}{c|c|c|c|c|c|c} \hline
        Instrument & Obs.Id. & Date & Exposure & Flux 0.3-10\,keV        & Index          & $n_H$\\
                   &         &  MJD &   ksec   & $10^{-12}$~erg\,cm$^{-2}$\,s$^{-1}$ &                & $10^{22}$cm$^{-2}$\\
        \hline 
        \cha      & 11831   &55425.039 &    10    &     $1.65\pm0.1$         &$1.3\pm0.25$    & $0.8\pm0.4$    \\
        \cha      & 14657   &56650.855 &    45    &     $1.55\pm0.05$        &$1.5\pm0.1$     & $1.2\pm0.2$    \\
        \cha      & 16560   &56655.036 &    28    &     $1.07\pm0.06$        &$1.65\pm0.2$    & $1.3\pm0.4$    \\
        \xmm     &0604700101&55065.695 &    20    &     $0.79\pm0.03$        &$1.55\pm0.1$    & $0.85\pm0.1$    \\
        \xmm     &0694390101&56301.594 &    100   &     $1.505\pm0.01$       &$1.57\pm0.04$   & $1.07\pm0.05$   \\ \hline
    \end{tabular}

\end{table*}

\subsubsection{XMM-Newton}
Two observations of \fgl  were taken  by \xmm in 2009 and 2013, see Table~\ref{tab:goodxrays}. The data were analysed with \xmm \texttt{Science Analysis Software} (v. 19.1.0). Known hot pixels and electronic noise were removed, and the data were filtered to exclude soft proton flares episodes. The spectrum was extracted from a $40''$-radius circle centred at the position of \fgl and the background was extracted from a nearby source-free region with a radius of  $80''$. The RMFs and ARFs were extracted using the \texttt{RMFGEN} and \texttt{ARFGEN} tools, respectively. Similar to the \cha data analysis we fit the derived spectra with the absorbed power-law model, see best-fit parameters in Table~\ref{tab:goodxrays}.

\subsubsection{Swift/XRT}
Publicly available \xrt data on \fgl have been taken between September 2009 and September 2019. The data were reprocessed with \texttt{xrtpipeline} v.0.13.6 as suggested by the \xrt team.\footnote{See \href{https://swift.gsfc.nasa.gov/analysis/xrt_swguide_v1_2.pdf}{\xrt data reduction guide}} Spectra were extracted with the \texttt{xselect} routine at the position of \fgl  using a $36''$ circle for source counts and an annulus also centered at the same coordinates with an inner (outer) radius of $60''$ ($300''$) for background counts.

\begin{figure}
    \centering
    \includegraphics[width=\columnwidth]{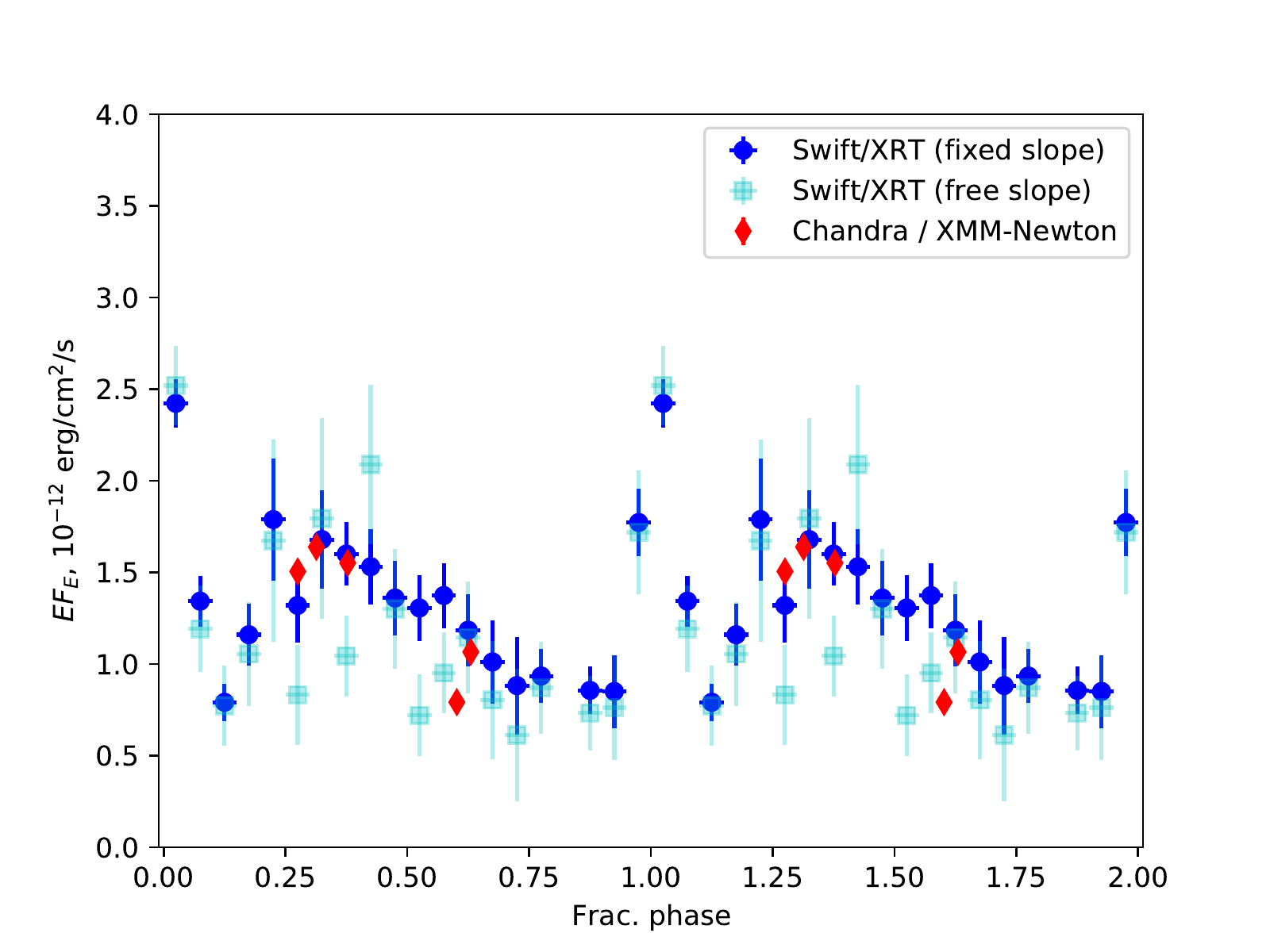}
    \caption{X-ray (0.3-10\,keV) orbital lightcurve of \fgl. The blue and cyan points are derived from \xrt data modelled with an absorbed power law model (free and fixed slope correspondingly). Red points stand for high quality \xmm and \cha observations. Data folded on a period of $16.5507$\,d (see section~\ref{sec:results}) , using $T_0 =  55403.3$.  See text for the details. }
    \label{fig:xray_lc}
\end{figure}

We analysed \xrt spectra on an observation-by-observation basis fitting the spectra with an absorbed power-law model. In order to minimize uncertainties connected to the poor statistics of the \xrt data at the initial step of our analysis we additionally fixed the $n_H$ and the power-law index of the model to values motivated by the higher quality \xmm and \cha observations ($n_H=10^{22}$~cm$^{-2}$, $\Gamma=1.5$). The blue points in Fig.~\ref{fig:xray_lc} correspond to the average (weighted with flux uncertainties) flux values of the observations performed at corresponding orbital phases. 

In addition to the approach described above we performed the analysis assuming only a fixed $n_H$ value and allowed the spectral slope to vary between the observations. The averaged flux values (weighted with flux uncertainties) in this approach are shown with cyan points in Fig.~\ref{fig:xray_lc}. 

We note a generally good agreement of the results derived by the different  approaches. However, at orbital phases $0.25-0.6$ the agreement is substantially worse than at other phases. This could indicate the possible fast (observation-to-observation) variations of the spectral index at these orbital phases. We argue that the \xrt data alone cannot provide sufficient sensitivity to track such variations. If the variations in the spectral index are confirmed by future high quality X-ray observations,  then the previously obtained \xrt results are indicative of fast X-ray flux/spectral index variability at phases $0.2-0.65$ and not the presence of a second peak in the X-ray light curve.

\subsection{{\it Fermi}-LAT observations}

The analysis of {\it Fermi}-LAT data was performed using Fermitools version 2.0.8 (released 20th January 2021). This was carried out using the latest Pass 8 reprocessed data (P8R3) \citep{atwood2013pass} from the SOURCE event class for 13 years worth of data from the region of 1FGL J1018.6-5856. For the binned likelihood analysis, all gamma-ray photons selected were within the energy range $0.1 - 500.0$ GeV and within a circular region of $15^{\circ}$ around the region-of-interest centred on 1FGL J1018.6-5856, with a maximum zenith angle of $90^{\circ}$. The spatial-spectral model used to perform the likelihood analysis included the Galactic and isotropic diffuse emission components and all known gamma-ray sources within $20^{\circ}$ of the ROI centre from the 4FGL catalogue \citep{4FGL}. In the 4FGL catalogue, 1FGL J1018.6-5856 uses the identifier 4FGL J1018.9-5856. In the likelihood fit of the total data set used, the normalization of all sources within $5^{\circ}$ of the ROI, the Galactic diffuse model and the isotropic diffuse model were free. The alpha and beta parameters of the LogParabola model of 1FGL J1018.6-5856 were also free during the likelihood analysis. The parameters of all other sources within the model were fixed. 

In order to investigate the periodicity of the source of interest, 1FGL J1018.6-5856, aperture photometry lightcurves were built for a range of energy ranges. These lightcurves used only events within $1^{\circ}$ of 1FGL J1018.6-5856. The lightcurves were built with a linear time binning where the bins had a width of $60$~s. The exposure for the time bins were calculated using the spectral model of 1FGL J1018.6-5856, as calculated during the initial likelihood analysis of the total selected data period. The advantage of using these aperture photometry lightcurves is that the rate data can be folded using any value for the period without any significant computational time and thus many period values were tested using the methods described in the Results.

\section{Results}
\label{sec:results}
\subsection{Orbital period}

\begin{figure}
    \centering
    \includegraphics[width=\columnwidth]{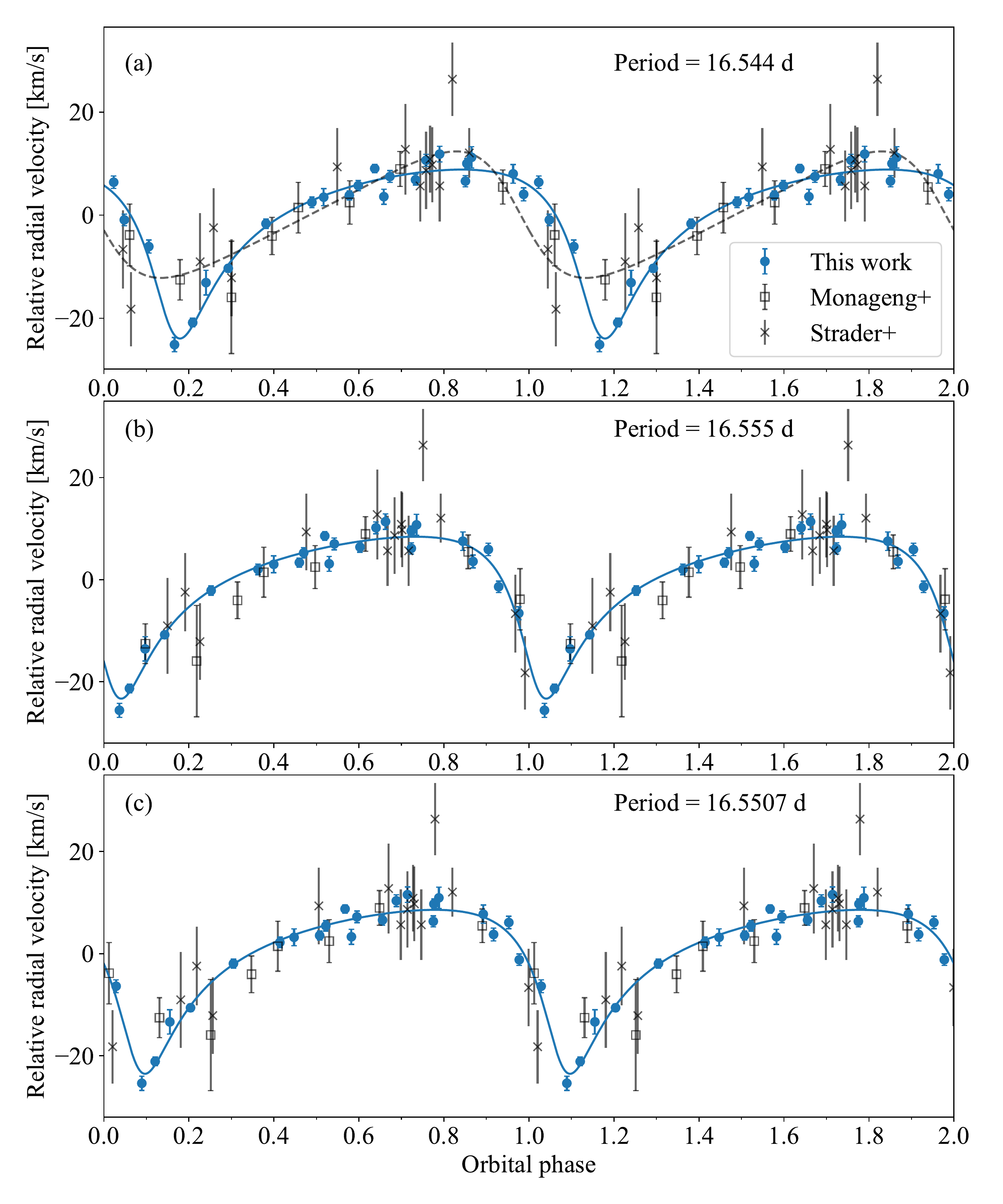}
    \caption{Relative radial velocties measured for \fgl\ folded on three different periods, using the ephemeris $T_0 =  55403.3$. For each plot the systemic velocity has been subtracted for comparison and the data is repeated over two orbits for clarity. In each plot the data reported in this paper are shown as filled blue circles, the SALT observations from \citet{monageng17} are shown as open squares, and the measurements from \citet{strader15} are shown as grey crosses. In the top panel the dashed grey line shows the fit found in \citet{monageng17}. The radial velocities are folded on the period reported in \citet{an15} (a), the period found by the fit to the new radial velocity measurement (b), and on period found from the new analysis of the {\it Fermi}-LAT data (c).}
    \label{fig:rv_plots}
\end{figure}

The radial velocities found from the current observations are compared to those reported in \citet{monageng17} in Fig.~\ref{fig:rv_plots}, where the systemic velocity has been subtracted for a direct comparison.  Panel (a) shows the best fit solution calculated for only the new data (solid cyan line), compared to the previous solution found \citep[dashed grey line;][]{monageng17}, if the period is held at 16.544\,d. The two solutions show a large difference between phases $\phi \approx 0.0 - 0.2$. If the orbital period is left as a free parameter, the best fit to the new data finds an orbital period of $16.555 \pm 0.019$\,d as shown in panel (b) of Fig.~\ref{fig:rv_plots}. The slightly longer orbital period appears to be more compatible with the previous observations.
In order to independently search for the orbital period we undertook a Lomb-Scargle, $\chi^2$ and auto-correlation analysis of the {\it Fermi}-LAT photons counts using $\sim$13\,years of observations.

\subsubsection{Lomb-Scarlge analysis}

\begin{figure}
    \centering
    \includegraphics[width=\columnwidth]{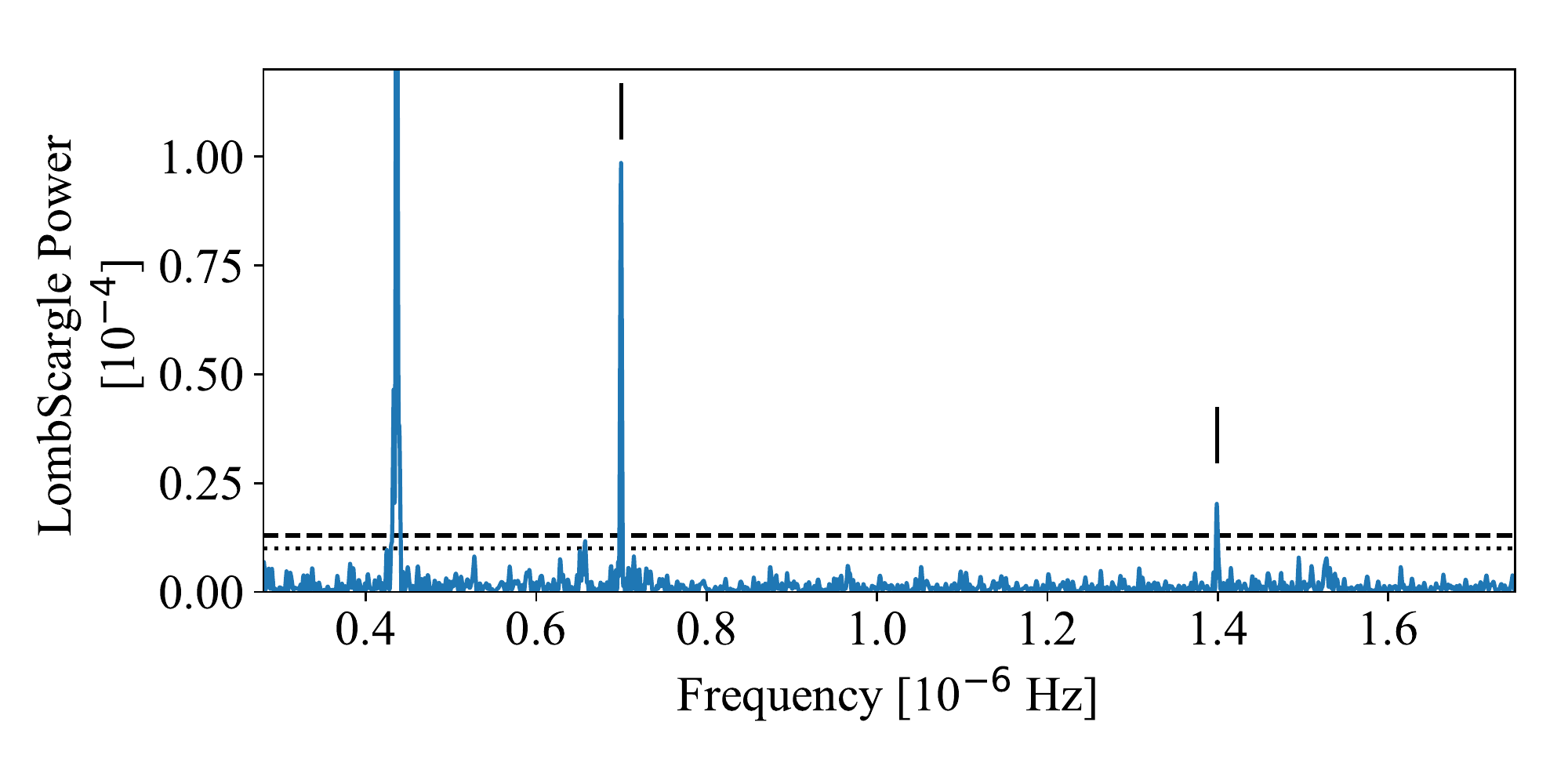}
    \caption{Lomb-Scargle periodogram for the {\it Fermi}-LAT photon count, between $\approx6.6 - 41$\,d. The peak frequency ($\approx 0.7$\,$\umu$Hz) and the first harmonic ($\approx 1.7\,\umu$Hz) are indicated. The $2\sigma$ and $3\sigma$ significant levels are indicated by dotted and dashed lines, respectively. }
    \label{fig:ls}
\end{figure}

A Lomb Scargle analysis was performed using the implementation in {\sc astropy} (v. 4.3.post1). The periodogram between $\approx6.6 - 41$\,d. is shown in Fig.~\ref{fig:ls}, where the binary period ($\approx 0.7$\,$\umu$Hz) and the first harmonic ($\approx 1.7\,\umu$Hz) are indicated. The peak at $\approx 0.44\,\umu$Hz is an artifact of the sampling. The $2\sigma$ and $3\sigma$ confidence levels have been estimated via a bootstrap method, using 10\,000 samplings. 

\begin{figure}
 \centering
 \includegraphics[width=\columnwidth]{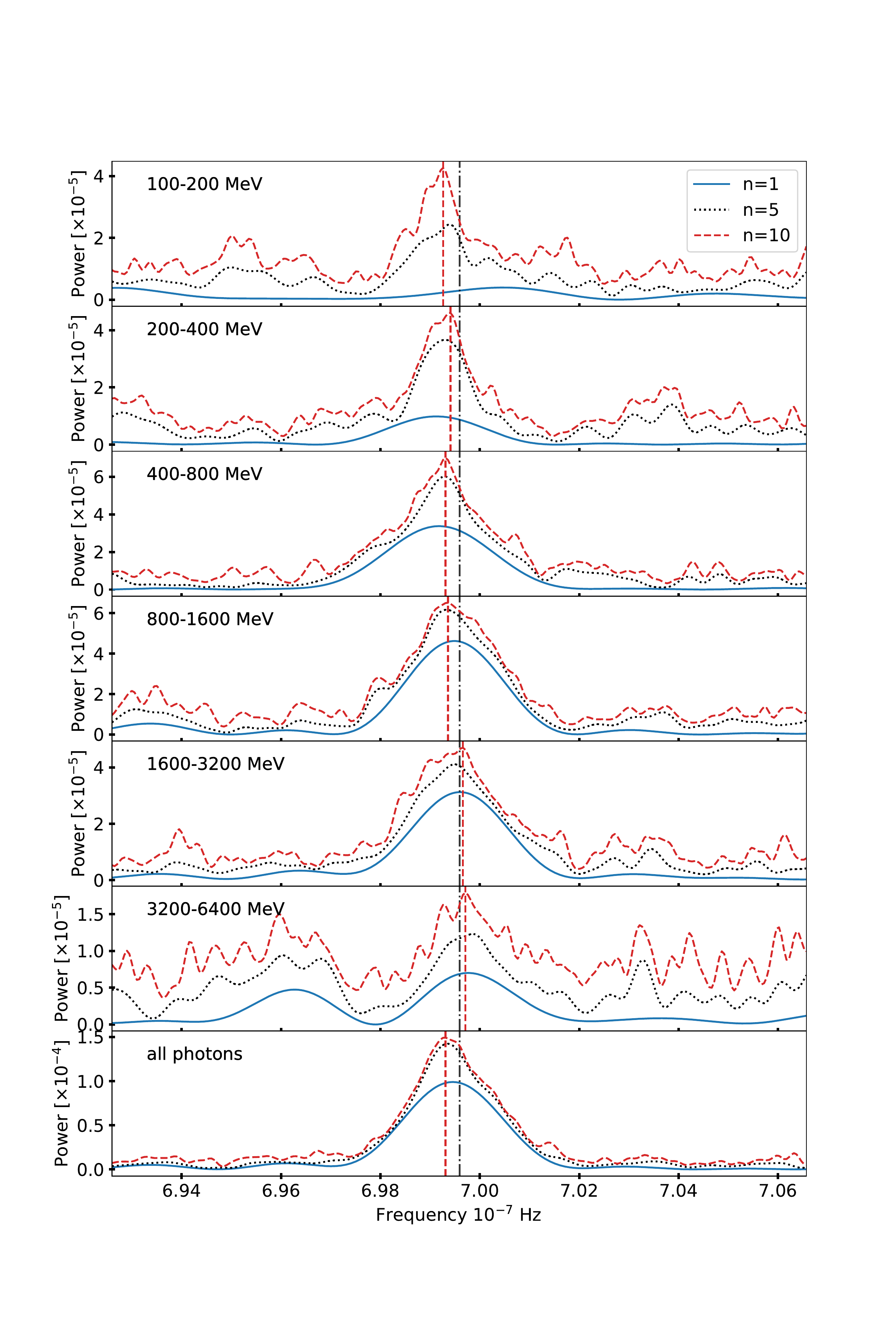}
 \caption{ Lomb Scargle periodogram of the {\it Fermi}-LAT data in different energy ranges, as indicated in the figure. The periodogram for 1 (blue solid line), 5 (dotted black line) and 10 (dashed red line) terms are shown. The peak for 10 terms is indicated by a vertical red dashed line. The vertical black dot-dash line shows the 16.544\,d period found by \citet{an15}. }
 \label{fig:ls_diff_energies}
\end{figure}

The classical Lomb Scargle method strictly searches for sinusoidal modulation. For non-sinusoidal modulation, a better fit can be found by considering a multi-term Fourier model  \citep{vanderplas18}. In order to better refine the peak frequency, periodograms were calculated around the peak associated with the $\sim16.5$\,d period, using $\Delta f = 5.0085\times10^{-11}$\,Hz frequency steps, for 1, 5 and 10 Fourier terms (1 term is the classical Lomb Scargle method).  The periodograms were calculated for the {\it Fermi}-LAT data in 6 different energy bands as well as for all photons (Fig.~\ref{fig:ls_diff_energies}). Apart from the $1\,600-3\,200$\,MeV and $3\,200-6\,400$\,MeV energy ranges, the periodograms peak at a slightly lower frequency (longer period) than previously found \citep{an15}. The periods at which each of the periodograms peak is given in Table~\ref{tab:ls_peaks}. For all photons, using 10 Fourier terms, the periodogram peaks at a period of $P=16.5507\pm0.0004$\,d. 

\begin{table}
 
{%
\newcommand{\mc}[3]{\multicolumn{#1}{#2}{#3}}
\begin{center}
\caption{The period (days) at which which each Lomb Scargle periodogram peaks for different number of Fourier terms.}
\label{tab:ls_peaks}
\begin{tabular}{lccc} \hline
Energy & \mc{3}{c}{Peak period (days)}\\
 & $n=1$ & $n=5$ & $n=10$\\ \hline
100-200 MeV & 16.5235 & 16.5484 & 16.5519\\
200-400 MeV & 16.5555 & 16.5507 & 16.5484\\
400-800 MeV & 16.5543 & 16.5507 & 16.5507\\
800-1600 MeV & 16.546 & 16.5507 & 16.5495\\
1600-3200 MeV & 16.5436 & 16.546 & 16.5424\\
3200-6400 MeV & 16.5401 & 16.5377 & 16.5413\\
All photons & 16.5472 & 16.5495 & 16.5507 \\ \hline
\end{tabular}
\end{center}
}%

\end{table}

\subsubsection{$\chi^2$ analysis}

\begin{figure}
 \centering
 \includegraphics[width=\columnwidth]{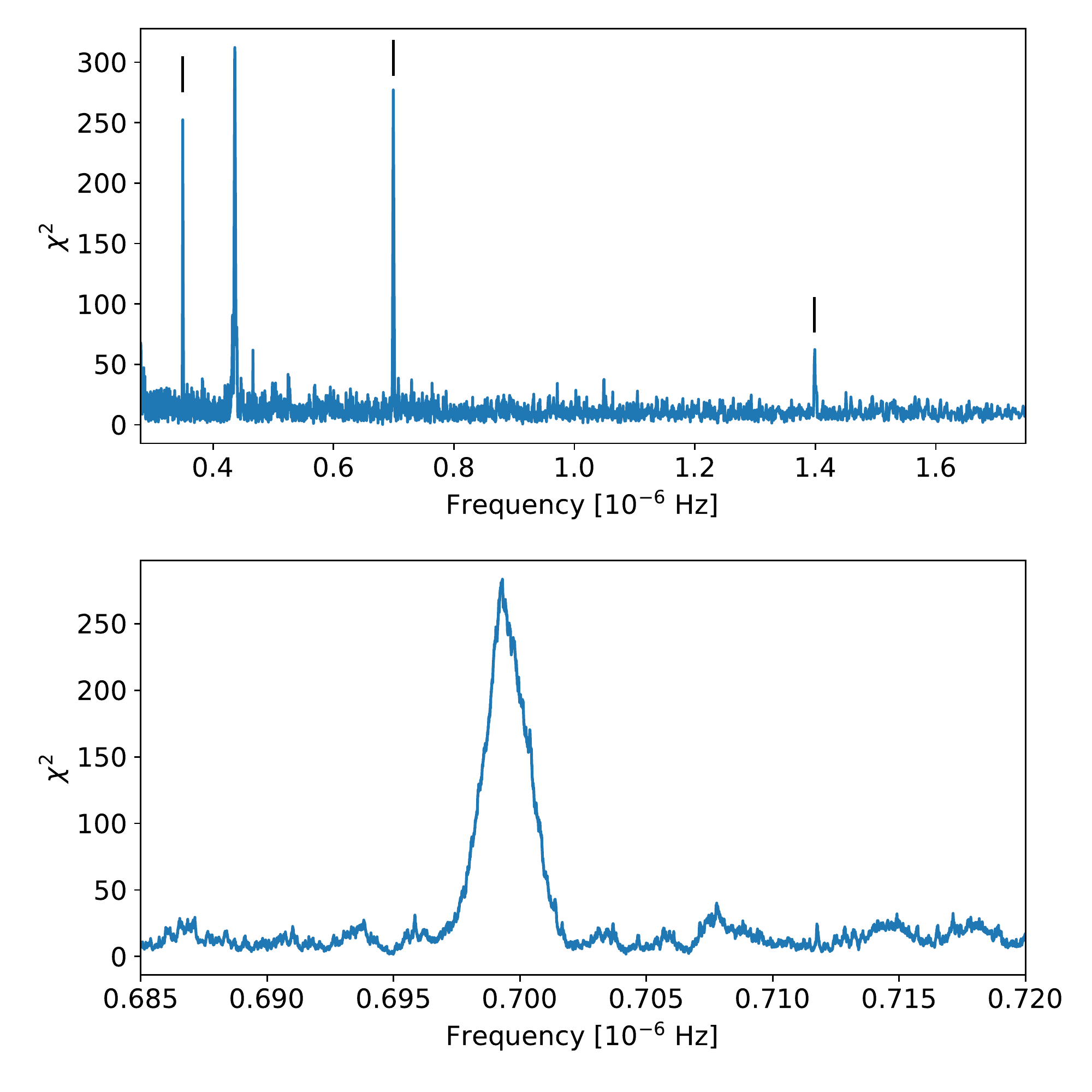}
 \caption{$\chi^2$ analysis of the {\it Fermi}-LAT data. Top: periodogram showing the period  ($\approx 0.7$\,$\umu$Hz) and two harmonics  ($\approx 0.35$\,$\umu$Hz and $\approx1.4$\,$\umu$Hz) of the observations. Bottom: region around the orbital period peak. }
 \label{fig:chi2}
\end{figure}

We also considered a $\chi^2$ analysis to search for the period. The {\it Fermi}-LAT data was folded on various test periods, and binned into 10 phase bins. A $\chi^2$ fit to a straight line was performed, to search for the largest deviation from this straight line fit (largest $\chi^2$ value). Fig.~\ref{fig:chi2} shows the $\chi^2$ plotted versus frequency. The peak at the orbital period is clearly visible ($0.699\,\umu$Hz) along with the two harmonics (top panel). Using a time step of $\Delta t = 5.0\times 10^{-5}$\,d, (bottom panel) the $\chi^2$ peaks at a period of 16.5507\,d.

\subsubsection{Logliklihood analysis}

\begin{figure}
 \centering
 \includegraphics[width=\columnwidth]{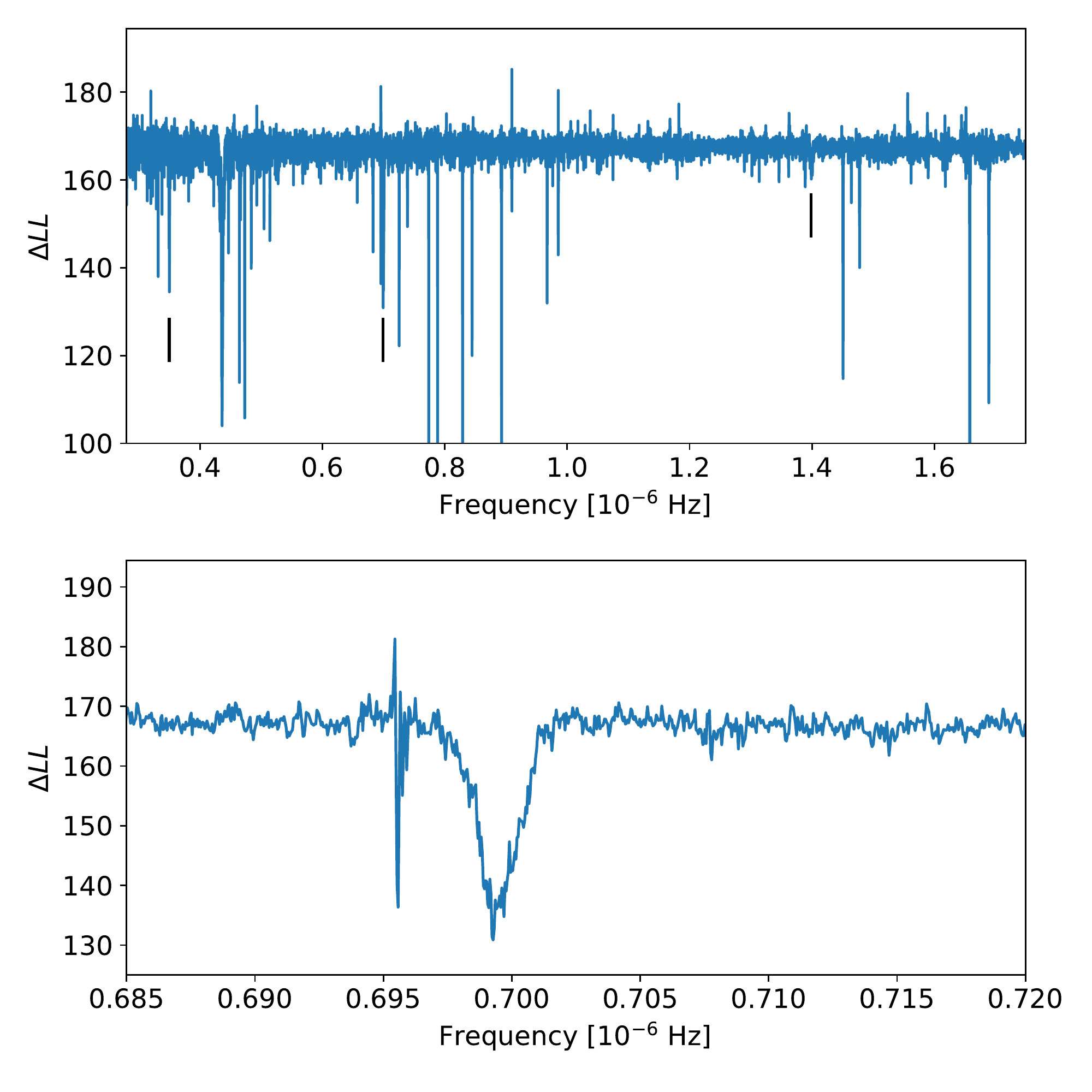}
 \caption{The results for autocorrelation analysis. Top: periodogram where the orbital period ($\approx 0.7$\,$\umu$Hz) and two harmonics  ($\approx 0.35$\,$\umu$Hz and $\approx1.4$\,$\umu$Hz) are indicated. Bottom: region around the orbital period peak.  }
 \label{fig:ll}
\end{figure}

Lastly we considered an auto-correlation analysis. The data was folded on test periods between 6.618\,d and 41.360\,d in time steps of $0.001$\,d, binned into 10 bins, and a cubic spline was fit to the folded data. The Pearson correlation coefficient was then calculated comparing this model to the original, unbinned, observed fluxes. This is shown in Fig.~\ref{fig:ll}, where lower values correspond to a better fit. The data shows a minimum at 16.5516\,d, while a Gaussian function fit to the profile finds a minimum at 16.5469\,d.

\subsubsection{Concluding remarks on the orbital period}
\begin{figure}
    \centering
    \includegraphics[width=\columnwidth]{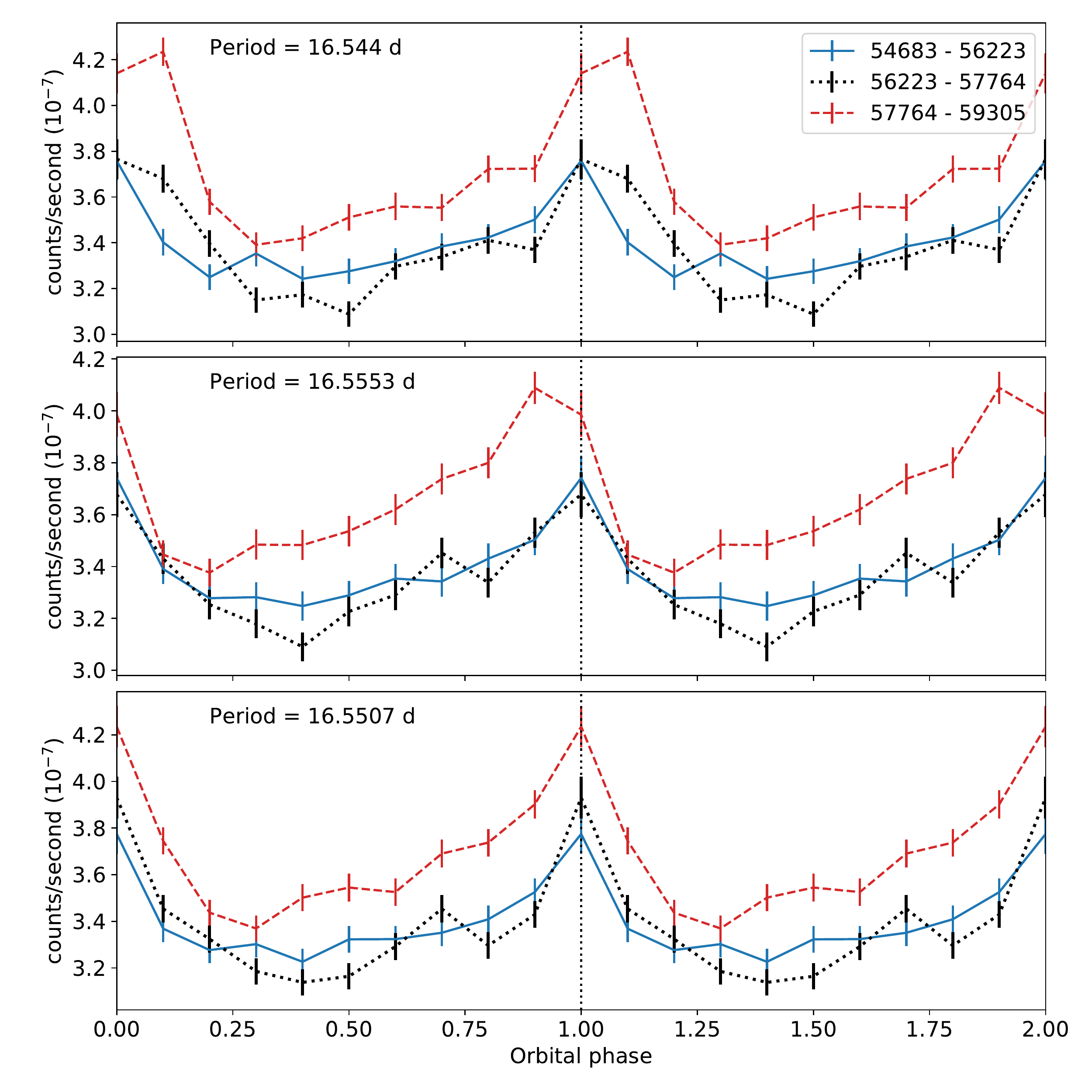}
    \caption{{\it Fermi}-LAT data, divided into three equal time sections and folded with the same orbital periods and ephemeris shown in Fig.~\ref{fig:rv_plots}. The data in binned into 10 bins, and repeated over two orbits for clarity. }
    \label{fig:comp_period}
\end{figure}

The different period analysis of the {\it Fermi}-LAT data all appear to suggest an orbital period that is slightly longer than previously found, and this is more consistent with the radial velocity curves. As a further confirmation, we divided the {\it Fermi}-LAT data into three equal time sections, and folded the data on the three periods shown in Fig.~\ref{fig:rv_plots}.  This shows (Fig.~\ref{fig:comp_period}) that using a period of $16.5507$\,d, the folded light curve consistently peaks at the phase 0, over the $\approx13$ years of observations.  We, therefore, adopt this as the orbital period.   The radial velocities, folded on this period are shown in panel (c) of Fig.~\ref{fig:rv_plots}, along with the best fit to this period. 

\begin{figure}
    \centering
    \includegraphics[width=\columnwidth]{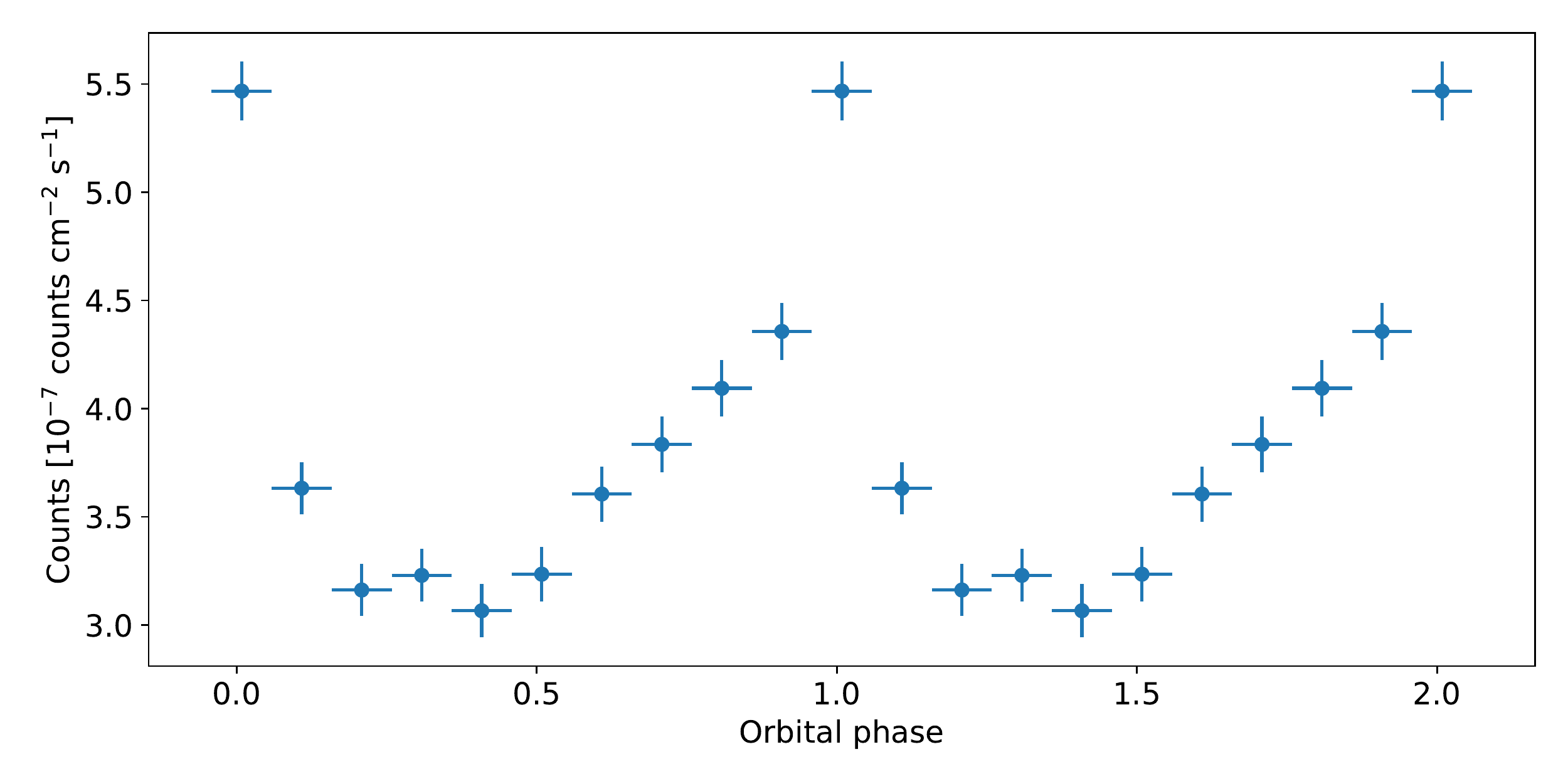}
    \caption{{\it Fermi}-LAT data binned into 10 orbital phases based on the periodicity of 16.5507 d, using an ephemeris of $T_0 =  55403.3$. The flux of each bin is calculated with a binned likelihood analysis of the selected events. Photon events selected were taken from an energy range in $0.1-500$~GeV.}
    \label{fig:folded_like}
\end{figure}

The orbital phase light curve calculated from 13 years of
observational data from {\it Fermi}-LAT, using a period of $16.5507$\,d, is shown in Fig. \ref{fig:folded_like}. This was done with a binned likelihood analysis using the output source model from the initial likelihood analysis of the total observational data used.

\subsection{Orbital parameters}

The orbital parameters derived from the new SALT observations are given in Table~\ref{tab:orbital_parameters}, where we show the results for all three orbital periods considered. The systemic velocities are shown as measured against the template, and when the velocity of the template is considered. We have estimated the average radial velocity of the template by fitting a Gaussian to the core of the H\,$\beta$, He\,{\sc i}\,$\lambda 4921$, and the He\,{\sc ii} $\lambda 4686, \lambda4542$ and $\lambda 5411$ lines. We find a difference in the velocities for the different lines as has previously been reported \citep[e.g.][]{strader15,monageng17}. The average velocity of the template was measured as $56.8 \pm 13.1$  km\,s$^{-1}$, where the error is taken as the standard deviation of the measurements.   The systematic velocity give in Table.~\ref{tab:orbital_parameters} is found by adding this value to that measured relative to the template. 

\begin{table*}
 \centering
 \caption{Orbital parameters of \fgl\ found from the fit to the radial velocities. The three solutions presented are for the period held at the previously published period, for the period left as a free parameter, and for the period held at that found by the Lomb Scargle and $\chi^2$ analysis. }
 \label{tab:orbital_parameters}
 \begin{center}

 {%
\newcommand{\mc}[3]{\multicolumn{#1}{#2}{#3}}
\begin{center}
\begin{tabular}{lr@{ $\pm$ }lr@{ $\pm$ }lr@{ $\pm$ }l} \hline
 & \mc{2}{c}{Held} & \mc{2}{c}{Free fit} & \mc{2}{c}{Held at new period (adopted)}\\ \hline
$T_p$ (time of periastron; MJD) & 57258.85 & 0.17 & 57257.81 & 1.79 & 57258.23 & 0.17\\
Orbital period (d) & \mc{2}{c}{16.5440} & 16.5553 & 0.0189 & \mc{2}{c}{16.5507}\\
Systemic velocity relative to template (km\,s$^{-1}$). & -1.78 & 0.42 & -1.31 & 0.85 & -1.50 & 0.41\\
Systemic velocity (km\,s$^{-1}$) & 55.0 & 13.1 & 55.5 & 13.2 & 55.3 & 13.1\\
Velocity semi-amplitude (km\,s$^{-1}$) & 16.39 & 0.90 & 15.87 & 1.00 & 16.07 & 0.87\\
eccentricity & 0.515 & 0.034 & 0.543 & 0.073 & 0.531 & 0.033\\
longitude of periastron (degree) & 153.5 & 5.0 & 149.8 & 9.1 & 151.2 & 5.1\\
Mass function (M$_\odot$) & 0.00475 & 0.00085 & 0.00406 & 0.00102 & 0.00432 & 0.00077 \\ \hline
 \end{tabular}
 \end{center}
}%
 \end{center}
\end{table*}

\section{Discussion}

\subsection{Mass of the compact object}

\begin{figure}
    \centering
    \includegraphics[width=\columnwidth]{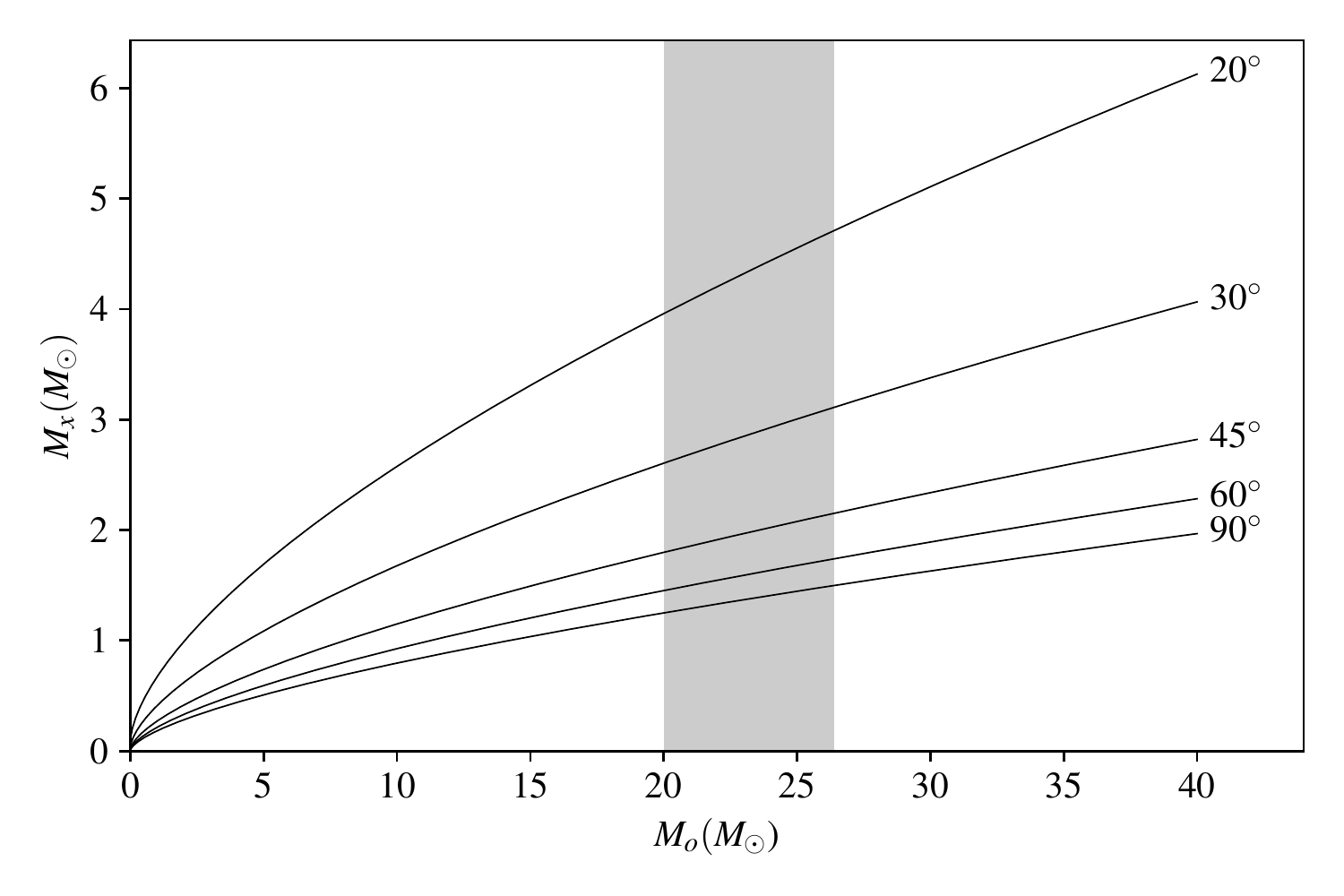}
    \caption{The constraints on the mass of the compact object in \fgl. The shaded region indicates the mass range of the companion star.}
    \label{fig:mass_function}
\end{figure}

The constraint on the mass of the compact object for a mass function  of $f =  0.00432\pm 0.00077$\,M$_\odot$ (for $P=16.5507$\,d) is shown in Fig.~\ref{fig:mass_function}. The mass function found from the  newer analysis is larger than previously found \citep[$f = 0.0027 \pm  0.0013$\,M$_\odot$;][]{monageng17}. Using a mass range of $20-26.4$\,\msun\ for the optical companion \citep{casares05,strader15,waisberg15,monageng17}, we can place a constraint on the inclination angle. For a neutron star mass of the 1.4\,\msun\ and a mass of 20\,\msun\ for the optical companion, the inclination would be $i=64\pm7^\circ$, but a solution is not possible for the higher mass of the optical companion.  For a upper-mass value for a neutron star of 2.0\,\msun, the inclination is $i=40\pm3\degr$ and $i = 49\pm4\degr$ for the lower and upper mass of the optical companion, respectively. A black hole compact object, assuming a mass of $\approx5\,\msun$ \citep[e.g.][]{strader15}, would require inclinations angles $\lesssim 16 - 19\degr$. These results show that if the compact object is in the mass range of a neutron star, it is observed at a high inclination angle.

\subsection{Binary orientation}

\begin{figure}
    \centering
    \includegraphics[width=\columnwidth]{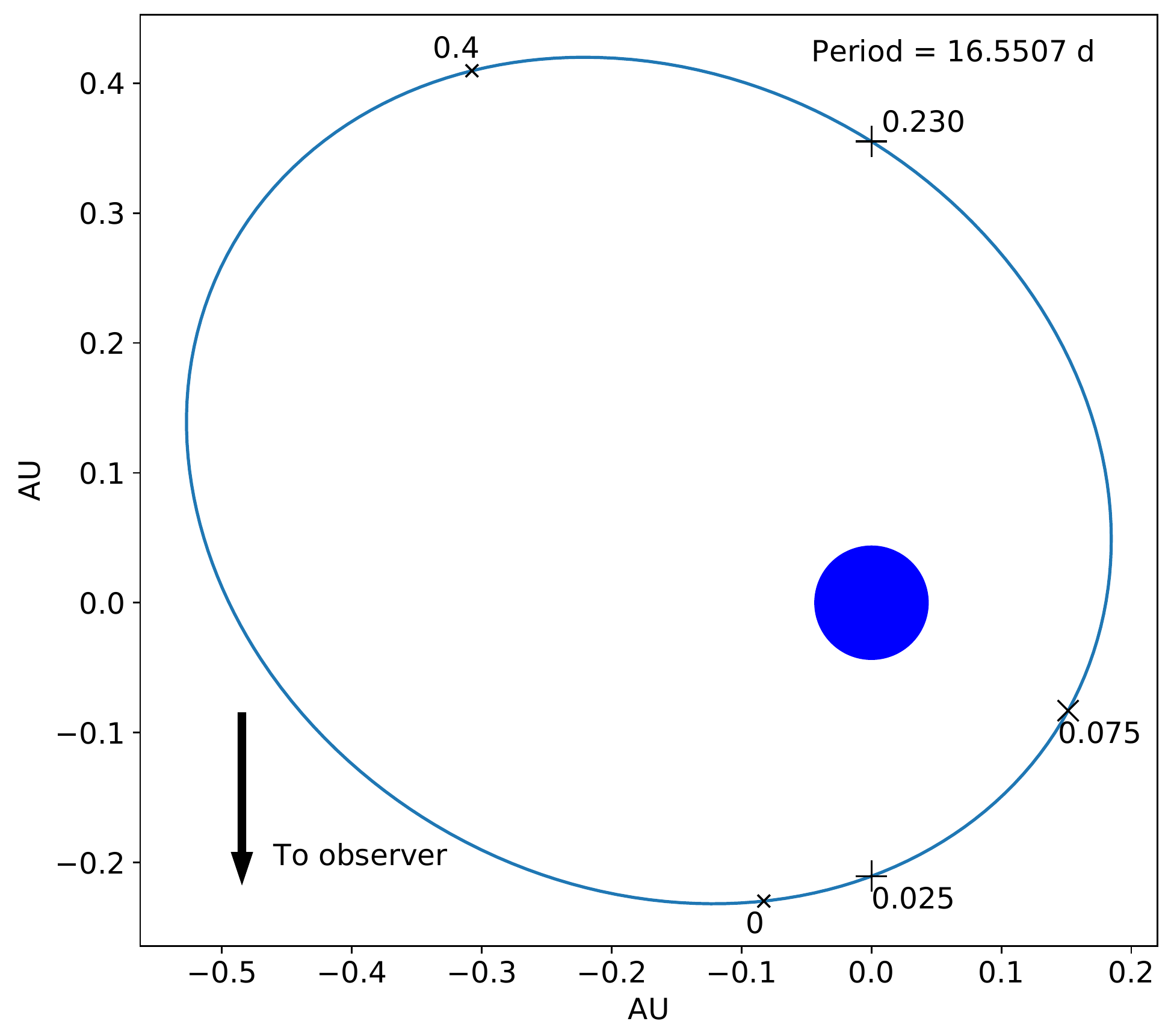}
    \caption{The binary solution using an orbital period of 16.5507\,d.  The phases of periastron, inferior conjunction and superior conjunction are marked. Phases 0 and 0.4 are also indicated on the plots. A mass of $M_x=2\,\msun$ is used for the the compact object and the optical companion has a mass of $M_\star = 22.9\,\msun$ and a radius of $R_\star = 9.3\,$R$_\odot$. }
    \label{fig:binary_orbit}
\end{figure}

The binary solution using an orbital period of 16.5507\,d is shown in Fig.~\ref{fig:binary_orbit}. While all three orbital periods give similar solutions, the critical phase of 0, corresponding to the peak in the {\it Fermi}-LAT light curve,  shifts depending on the assumed orbital period. Adopting the period of 16.5507\,d, the position of phase 0 is close to inferior conjunction, while periastron is at 0.075. 

\subsection{Non-thermal emission}

Multiwavelength orbital light curves of \fgl,  as well as the presented orbital solution for this system,  allow us to propose a scenario which explains the observed high energy (keV-TeV) emission. We argue that the system can be described by a model similar to that proposed by~\citet{psrb_2017,psrb_2021} to explain the emission produced from PSR~B1259-63 close to periastron. Namely, we propose that in the \fgl system at orbital phases close to the inferior conjunction ($\phi=0.025$) the interaction of the compact object (presumably a pulsar) and stellar outflows lead to a formation of the cone-like surface oriented towards the observer. X-ray to GeV emission could be explained as synchrotron emission from the relativistic electrons of the pulsar wind flowing along the surface of the cone. TeV emission corresponds to the IC emission from the same population of electrons. 
As the pulsar approaches periastron ($\phi=0.075$) the cone is not oriented towards the observer anymore which leads to the rapid drop of the emission level in all energy bands. Even further from the periastron, e.g. close to superior conjunction ($\phi=0.23$), the density of the stellar outflow significantly decreases; the outflow becomes more heterogeneous and clumpier which results in destruction of the smooth stellar/pulsar wind interaction cone-like surface. We suggest that at these orbital phases multiple randomly oriented stellar wind clumps/pulsar wind interaction surfaces are formed. Rather variable X-ray-to-GeV synchrotron emission could be formed at these surfaces. The IC-originated TeV emission at these orbital phases remains at relatively low levels due to the substantially decreased (in comparison to periastron) density of the soft photons.

\section{Conclusions}

We have undertaken SALT/HRS observations of \fgl\ finding an improved orbital solution. The source has an eccentricity of $e\approx 0.53$ and an orientation (longitude of periastron $\approx 151$\degr) which is different to what was found from previous observations. The larger mass function found also shows that if the compact object is in the mass range of a neutron star, the system must be observed at a high inclination angle.  This solution is found using an updated orbital period of 16.5507\,d established from an analysis of approximately 13 years of {\it Fermi}-LAT observations.  The peak in the {\it Fermi}-LAT light curve remains around orbital phase $\phi \approx 0$, which  is around inferior conjunction ($\phi_{ic}=0.025$) and periastron ($\phi_p = 0.075$). The phase of the second maximum in the X-ray light curve ($\phi = 0.2 - 0.65$) lies between superior conjunction and the apastron. Our reanalysis of X-ray data suggests that this may be an artefact of rapid variation in the X-ray slope, but further X-ray observations are required.

\section*{Acknowledgements}

The authors are grateful to P.A. Charles for valuable discussions.

All of the optical observations reported in this paper were obtained with the Southern African Large Telescope (SALT) under programmes 2018-1-MLT-001 and 2020-1-DDT-003 (PI: B van Soelen).
This work has made use of data from the European Space Agency (ESA) mission
{\it Gaia} (\url{https://www.cosmos.esa.int/gaia}), processed by the {\it Gaia}
Data Processing and Analysis Consortium (DPAC,
\url{https://www.cosmos.esa.int/web/gaia/dpac/consortium}). Funding for the DPAC
has been provided by national institutions, in particular the institutions
participating in the {\it Gaia} Multilateral Agreement.
We acknowledge the use of public data from the Swift data archive and thank the entire Swift team for accepting and planning Target-of-Opportunity requests. This work made
use of data supplied by the UK Swift Science Data Centre at the University of Leicester.
The authors wish to acknowledge the Irish Centre for High-End Computing (ICHEC) for the provision of computational facilities for the purpose of the Fermi data analysis.
BvS acknowledges
that this work was supported by the Department of Science and
Technology and the National Research Foundation of South Africa
through a block grant to the South African Gamma-Ray Astronomy
Consortium.
%
SMK and MC acknowledges that this work was supported by ESA PRODEX grant 4000120711.
 The authors acknowledge support by the
state of Baden-W\"urttemberg through bwHPC. DM work was supported by DLR through the grant 50OR2104.

\section*{Data Availability}

The data underlying this article will be shared on reasonable request to the corresponding author.



\bibliographystyle{mnras}
\bibliography{binary} 








\bsp	
\label{lastpage}
\end{document}